\documentclass[aps,prb,reprint,superscriptaddress]{revtex4-2}

\usepackage{graphicx, amsmath, amssymb, xspace, braket, bm}
\usepackage[version=4]{mhchem}
\usepackage{color}
\usepackage{lineno,hyperref}

\newcommand{\sub}[1]{\ensuremath{_{\textrm{#1}}}}
\newcommand{\SSO}{\ce{Sr3SnO}\xspace}
\newcommand{\SxSO}{\ce{Sr_{3-x}SnO}\xspace}
\newcommand{\APO}{\ce{\textit{A}3\textit{B}O}\xspace}

\newcommand{\Tc}{\ensuremath{T\sub{c}}\xspace}
\newcommand{\rev}[1]{#1}

\begin{document}

% Use the \preprint command to place your local institutional report
% number in the upper righthand corner of the title page in preprint mode.
% Multiple \preprint commands are allowed.
% Use the 'preprintnumbers' class option to override journal defaults
% to display numbers if necessary
%\preprint{}

%Title of paper
\title{Penetration depth and gap structure in the antiperovskite\\oxide superconductor \SxSO revealed by \boldmath$\mu$SR}

% repeat the \author .. \affiliation  etc. as needed
% \email, \thanks, \homepage, \altaffiliation all apply to the current
% author. Explanatory text should go in the []'s, actual e-mail
% address or url should go in the {}'s for \email and \homepage.
% Please use the appropriate macro foreach each type of information

% \affiliation command applies to all authors since the last
% \affiliation command. The \affiliation command should follow the
% other information
% \affiliation can be followed by \email, \homepage, \thanks as well.
\author{Atsutoshi Ikeda}
\thanks{These authors contributed equally}
\email[]{a.ikeda@scphys.kyoto-u.ac.jp}
\affiliation{Department of Physics, Kyoto University, Kyoto 606-8502, Japan}
%\homepage[]{Your web page}

\author{Zurab Guguchia}
\thanks{These authors contributed equally}
\email[]{zurab.guguchia@psi.ch}
\affiliation{Laboratory for Muon Spin Spectroscopy, Paul Scherrer Institute, 5232 Villigen, Switzerland}

\author{Mohamed Oudah}
\affiliation{Stewart Blusson Quantum Matter Institute, University of British Columbia, Vancouver BC, V6T 1Z4 Canada}
\affiliation{Department of Physics, Kyoto University, Kyoto 606-8502, Japan}

\author{Shun Koibuchi}
\affiliation{Department of Physics, Kyoto University, Kyoto 606-8502, Japan}

\author{Shingo Yonezawa}
\affiliation{Department of Physics, Kyoto University, Kyoto 606-8502, Japan}

\author{Debarchan Das}
\author{Toni Shiroka}
\author{Hubertus Luetkens}
\affiliation{Laboratory for Muon Spin Spectroscopy, Paul Scherrer Institute, 5232 Villigen, Switzerland}

\author{Yoshiteru Maeno}
\affiliation{Department of Physics, Kyoto University, Kyoto 606-8502, Japan}

%Collaboration name if desired (requires use of superscriptaddress
%option in \documentclass). \noaffiliation is required (may also be
%used with the \author command).
%\collaboration can be followed by \email, \homepage, \thanks as well.
%\collaboration{}
%\noaffiliation

\date{\today}

\begin{abstract}
% insert abstract here
We report a $\mu$SR study on the antiperovskite oxide superconductor \SxSO.
With transverse-field $\mu$SR, we observed the increase of the muon relaxation rate 
upon cooling below the superconducting transition temperature $\Tc=5.4$~K, evidencing bulk superconductivity.
The exponential temperature dependence of the relaxation rate $\sigma$ at low temperatures suggests a fully gapped superconducting state.
We evaluated the zero-temperature penetration depth $\lambda(0)\propto1/\sqrt{\sigma(0)}$ to be around 320--\rev{1020}~nm.
\rev{Such a large} value is consistent with the picture of a doped Dirac semimetal.
Moreover, we revealed that the ratio $\Tc/\lambda(0)^{-2}$ is larger than those of ordinary superconductors 
and is comparable to those of unconventional superconductors.
The relatively high \Tc for small carrier density may hint \rev{at} an unconventional pairing mechanism beyond the ordinary phonon-mediated pairing.
In addition, \rev{zero-field $\mu$SR did not provide evidence of} broken time-reversal symmetry in the superconducting state.
These features are consistent with the theoretically proposed topological superconducting state in \SxSO, 
as well as with $s$-wave superconductivity.
\end{abstract}

% insert suggested keywords - APS authors don't need to do this
%\keywords{}

%\maketitle must follow title, authors, abstract, and keywords
\maketitle

% body of paper here - Use proper section commands
% References should be done using the \cite, \ref, and \label commands

\section{Introduction\label{introduction}}
% Put \label in argument of \section for cross-referencing
%\section{\label{}}

% If in two-column mode, this environment will change to single-column
% format so that long equations can be displayed. Use
% sparingly.
%\begin{widetext}
% put long equation here
%\end{widetext}

% figures should be put into the text as floats.
% Use the graphics or graphicx packages (distributed with LaTeX2e)
% and the \includegraphics macro defined in those packages.
% See the LaTeX Graphics Companion by Michel Goosens, Sebastian Rahtz,
% and Frank Mittelbach for instance.
%
% Here is an example of the general form of a figure:
% Fill in the caption in the braces of the \caption{} command. Put the label
% that you will use with \ref{} command in the braces of the \label{} command.
% Use the figure* environment if the figure should span across the
% entire page. There is no need to do explicit centering.

% \begin{figure}
% \includegraphics{}%
% \caption{\label{}}
% \end{figure}

% Surround figure environment with turnpage environment for landscape
% figure
% \begin{turnpage}
% \begin{figure}
% \includegraphics{}%
% \caption{\label{}}
% \end{figure}
% \end{turnpage}

Antiperovskite (inverse perovskite) oxides \APO are the materials crystallizing in the same structure 
as the ordinary perovskite oxides but with the reversed positions of the metal and oxygen~\cite{widera1980ubergangsformen}:
in antiperovskite oxides, oxygen is at the center of \ce{O$A$3} octahedra.
When $A$ is an alkaline-earth element and $B$ is a group 14 element, one can expect the ionic configuration \ce{($A$^{2+})3$B$^{4-}O^{2-}}
with a metallic anion for $B$ such as \ce{Sn^{4-}} or \ce{Pb^{4-}}, which is rare in oxides.
Indeed, this unusual metallic anion is directly observed in \SSO by the recent studies of
M\"{o}ssbauer spectroscopy~\cite{Oudah2019SrDeficiency, Ikeda2019Moessbauer}.

There \rev{has} been a number of investigations toward clarifying their peculiar electronic band 
structure~\cite{kariyado2011three, kariyado2012low, hsieh2014topological, Nuss2015tilting, 
Okamoto2016thermoelectric, oudah2016superconductivity, Obata2017Ca3PbOARPES, Kariyado2017Ba3SnO, 
Ikeda2018bandstructure, Niklas2018Sr3-xSnO, Suetsugu2018Sr3PbO, Kawakami2018TopoSC, Kitagawa2018Sr3-xSnO, 
Oudah2019SrDeficiency, Kariyado2019pseudoLandauLevels, Obata2019Ca3PbO, Ikeda2019Moessbauer}.
Theoretical calculations suggest a band inversion between the valence $B$-$p$ and conduction $A$-$d$ bands in some antiperovskite oxides 
containing heavy elements such as \ce{Ca3PbO} and \ce{Ba3SnO}.
Due to this band inversion, these materials belong to Dirac semimetals or topological crystalline insulators, 
with slightly gapped Dirac cones in the vicinity of the Fermi level of the bulk electronic 
states~\cite{kariyado2011three, kariyado2012low, hsieh2014topological, Kariyado2017Ba3SnO}.
Especially, \SSO is suggested to be in the vicinity of \rev{a} topological transition 
with barely inverted bands~\cite{hsieh2014topological, Ikeda2018bandstructure}.
\rev{Signatures} of Dirac electrons \rev{have} been observed experimentally in various antiperovskite oxides including \ce{Ca3PbO}~\cite{Obata2017Ca3PbOARPES}, 
\ce{Sr3PbO}~\cite{Suetsugu2018Sr3PbO}, and \SSO~\cite{Kitagawa2018Sr3-xSnO}.

Recently, some of the present authors discovered that \SxSO is the first superconductor 
among antiperovskite oxides~\cite{oudah2016superconductivity}.
Theoretical analyses revealed that the inverted band structure in \SSO can lead to a topological 
superconductivity with a high winding number originating from electrons with \rev{a} total angular momentum $J=3/2$~\cite{Kawakami2018TopoSC}.
In this theory, besides the ordinary $s$-wave superconductivity, 
unconventional superconductivities with a full gap and a point-nodal gap are suggested.

For investigations of the symmetry of the superconducting order parameter, the muon spin relaxation/rotation ($\mu$SR) technique is a powerful tool.
The London penetration depth $\lambda$ can be evaluated from the muon spin 
depolarization rate $\sigma$, measured in the vortex state of a type-II superconductor.
We can deduce the structure of the superconducting gap from the temperature dependence of $\lambda$ 
at temperatures much below the superconducting critical temperature \Tc:
In fully gapped superconductors, $\Delta\lambda^{-2}(T) = \lambda^{-2}(0) - \lambda^{-2}(T)$ exhibits an exponential temperature $T$ dependence; 
whereas in nodal superconductors, this quantity exhibits a power-law temperature dependence at low temperature~\cite{Guguchia2017MoTe2}.
Moreover, time-reversal-symmetry breaking of superconductivity can be detected 
by a change in the zero-field muon spin depolarization rate $\sigma\sub{zf}$ \rev{below} \Tc measured in zero magnetic field~\cite{Luke1998Sr2RuO4}.

In this paper, we report $\mu$SR results on the antiperovskite oxide superconductor \SxSO.
We \rev{observe} a clear increase of the transverse-field muon spin depolarization rate $\sigma\sub{tf}$ below \Tc,
which exhibits exponential behavior at low temperatures, 
indicating the absence of nodes in the superconducting gap.
The deduced London penetration depth for $T\to0$ is 320--1020~nm and the ratio 
$\Tc/\lambda(0)^{-2}$ is comparable to those of high-temperature superconductors.
This fact possibly indicates an unconventional pairing mechanism.
We also performed $\mu$SR \rev{at} zero field and did not detect breaking of the time-reversal symmetry.
These results are consistent with the theoretically \rev{proposed} unconventional superconductivity 
belonging to the same symmetry as the Balian-Werthamer (BW) state~\cite{Balian1963BWstate}, 
as well as with the ordinary $s$-wave superconductivity~\cite{Kawakami2018TopoSC}.

\section{Experiment\label{experiment}}
\subsection{Sample Preparation \& Characterization}
Polycrystalline samples of \SxSO were prepared from Sr (Sigma-Aldrich, 99.99\%) and SnO (Sigma-Aldrich, 99.99\%).
Three samples were used for the $\mu$SR studies: Sample A for the measurements down to 1.5~K and Samples B and C to 0.26~K\@.
They were synthesized in the procedure same as Method B described in Ref.~\cite{Niklas2018Sr3-xSnO}.
We confirmed that the samples are dominated by \SxSO~\cite{Nuss2015tilting} with the lattice parameters of 
$a=0.51429(4)$~nm in Sample A, $a=0.51435(3)$~nm in Sample B, and $a=0.51421(4)$~nm in Sample C\@.
These lattice parameter values were extracted from the powder x-ray diffraction (PXRD) patterns, 
as described in our previous report~\cite{Oudah2019SrDeficiency}.
Although it is not easy to determine the actual composition of the samples accurately, we expect that $x$ is close to 0.5 
because our previous study shows that the average Sr/Sn ratio is roughly 2.5 based on energy-dispersive x-ray spectroscopy~\cite{Niklas2018Sr3-xSnO}.
The sample for the specific-heat measurement (Sample D) was prepared from the stoichiometric ratio of Sr (Furuuchi, 99.9\%) 
and SnO (Furuuchi, 99.9\%) in the procedure same as Method A described in Ref.~\cite{Niklas2018Sr3-xSnO}.

\subsection{Magnetic susceptibility and specific heat}
Direct-current (DC) magnetization was measured with a commercial magnetometer using the 
superconducting quantum interference device (Quantum Design, MPMS-XL)\@.
We used powder samples that were sealed in plastic capsules inside an argon-filled glove box.
Alternating-current (AC) magnetic susceptibility was measured with a lock-in amplifier (Stanford Research Systems, SR830) 
and using an adiabatic demagnetization refrigerator on a commercial apparatus (Quantum Design, PPMS)~\cite{Yonezawa2015ADR}.
Heat capacity of a sintered chunk was measured using a \ce{^3He} refrigerator on a commercial apparatus (Quantum Design, PPMS).

\begin{figure}
\includegraphics[width=\linewidth]{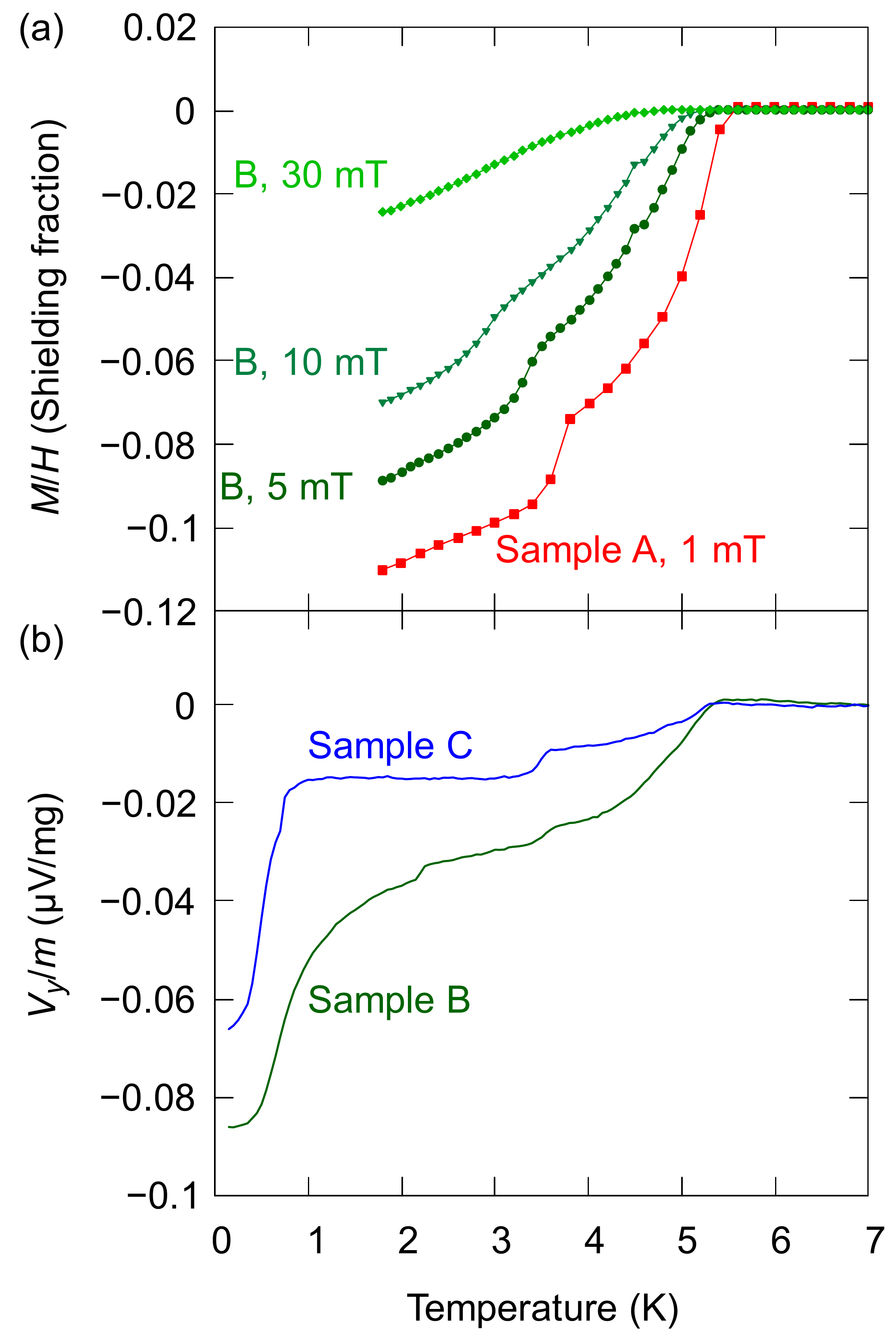}
\caption{(a) DC magnetic susceptibility of Sample A measured \rev{at} 1~mT and that of Sample B \rev{at} 5, 10, and 30~mT\@.
Both samples exhibit strong diamagnetism below $\Tc\simeq5.4$~K of \SxSO with the shielding fractions of around 10\% at 2~K\@.
The kink at 3.7~K originates from superconductivity of Sn impurity.
(b) Voltage signal detected with a mutual inductance coil normalized by the sample mass 
(corresponding to the real part of the AC susceptibility) of Samples B and C\@.
For both sets of data, the measurements were performed under an AC field with the amplitude of 16~$\mu$T and the frequency of 3011~Hz.
Both samples exhibit superconducting transition at $\Tc\simeq5.3$~K and 0.8~K\@.
Sample C exhibits a weaker 5-K transition but a sharper 0.8-K transition than Sample B\@.}
\label{fig: Susceptibility}
\end{figure}

\subsection{\boldmath$\mu$SR}
For $\mu$SR, we used pellets with a diameter of 10~mm.
Measurements on Sample A were performed at the GPS spectrometer ($\pi$M3 beamline) at the Paul Scherrer Institute (PSI) down to 1.5~K\@.
The pellet was sealed in a polyethylene bag with a thickness of 0.1~mm under argon atmosphere in order to avoid direct contact to air.
Measurements on Samples B and C were carried out at the Dolly spectrometer ($\pi$E1 beamline) at PSI down to 0.26~K\@.
Sample pellets were placed on 25-$\mu$m-thick Cu foils with grease (Apiezon, N Grease) to achieve a good thermal contact
and sealed with a polyimide film (Du Pont, Kapton) under nitrogen atmosphere.
The pellets of Samples B and C were mounted on the cryostat using a portable glovebox with continuous flow of nitrogen.
The typical integration time of $\mu$SR was 1.5 h for each temperature and field condition.

\section{Results\label{results}}
\begin{figure}
\includegraphics[width=\linewidth]{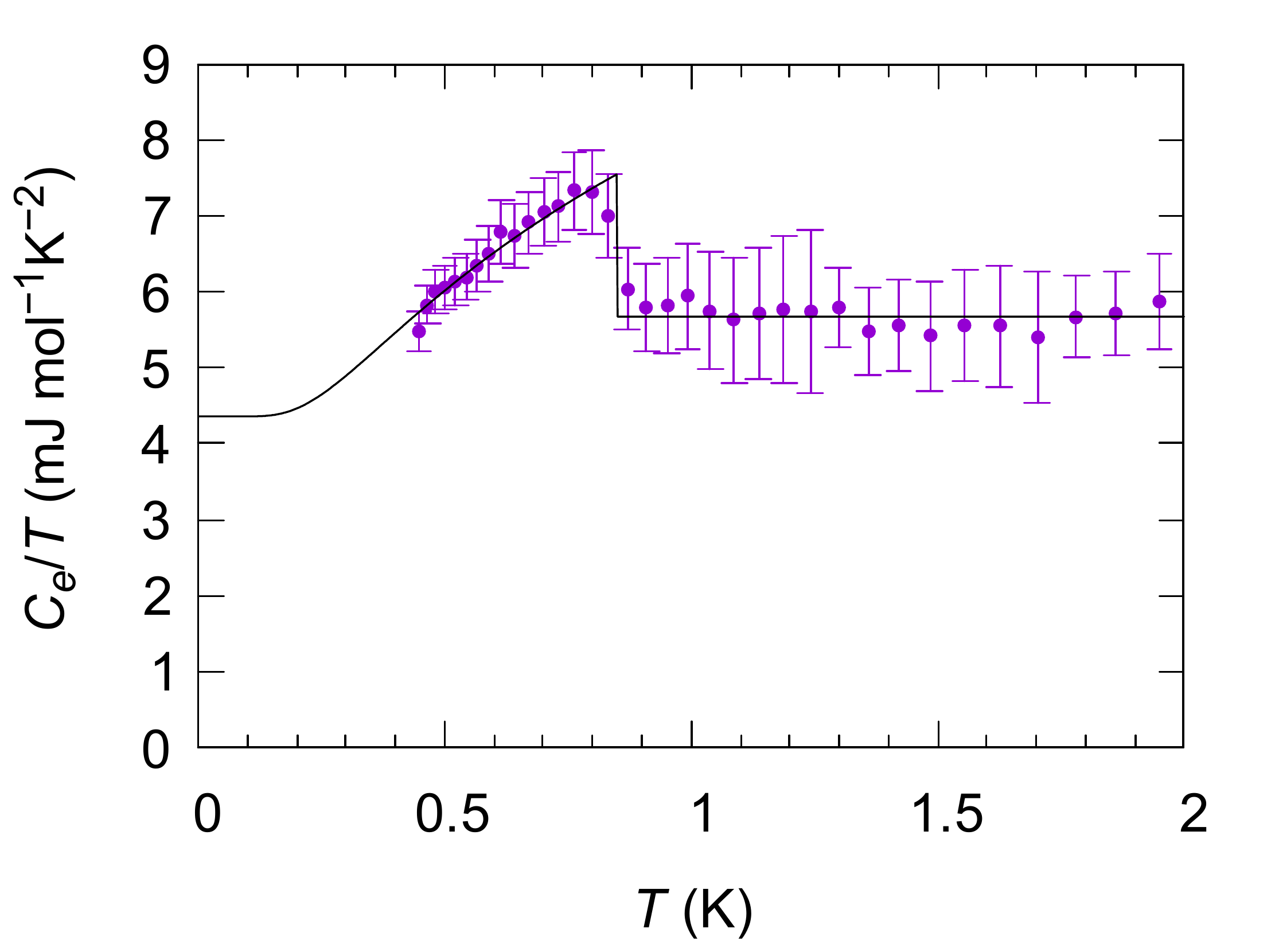}
\caption{Electronic specific heat divided by temperature $C_e/T$.
We observed a clear superconducting transition at 0.85~K with a volume fraction of 23\%, evidencing the bulk superconductivity.
The solid curve represent the fitting with the BCS theory.}
\label{fig: CPvsT}
\end{figure}

\subsection{Magnetic susceptibility and specific heat}
First, we show the DC magnetic susceptibility of the \SxSO samples in Fig.~\ref{fig: Susceptibility}.
All samples exhibit superconductivity at around 5.4~K\@.
We call this primary superconducting phase the 5-K phase.
The kink at 3.7~K is attributable to the superconductivity of Sn impurity, 
which was barely seen in PXRD patterns.
For Sample A, the superconducting volume fraction evaluated from the DC susceptibility 
without the demagnetization correction reaches 11\% at 1.8~K\@.
In the case of Sample B, the volume fraction is 9\% \rev{at} 5~mT and the fraction decreases with increasing the field.
Considering that the fields applied during the measurements of Sample B are stronger, 
we expect that the superconducting volume fraction of the 5-K phase is almost the same in both samples.
The volume fraction smaller than 100\% is attributable to the phase separation into multiple compositions 
with different amount of Sr deficiency~\cite{Kitagawa2018Sr3-xSnO, Oudah2019SrDeficiency}.
The onset of the transition decreases to 4.9~K \rev{at} 30~mT\@.
Figure~\ref{fig: Susceptibility}(b) shows the AC susceptibility down to 0.2~K of Samples B and C measured \rev{at} zero DC field.
Both samples exhibit additional superconducting transition at around 0.8~K as reported~\cite{oudah2016superconductivity}.
We hereafter call this second superconducting phase the 0.8-K phase.
The origin of the 0.8-K phase is not clear yet, 
but the separation into two phases with slightly different amount of 
the Sr deficiency may cause such splitting of the transitions~\cite{Oudah2019SrDeficiency}.
From these results, we infer that Sample B contains more 5-K phase than the 0.8-K phase, whereas Sample C contains more 0.8-K phase.
Sample D has a diamagnetic ratio between 5-K and 0.8-K phases similar to Sample C, or Sample D is dominated by the 0.8-K phase.
Thus, comparing the $\mu$SR results of these samples, we may unveil the difference in the superconducting properties of the 5-K and 0.8-K phases.

\begin{figure}
\includegraphics[width=\linewidth]{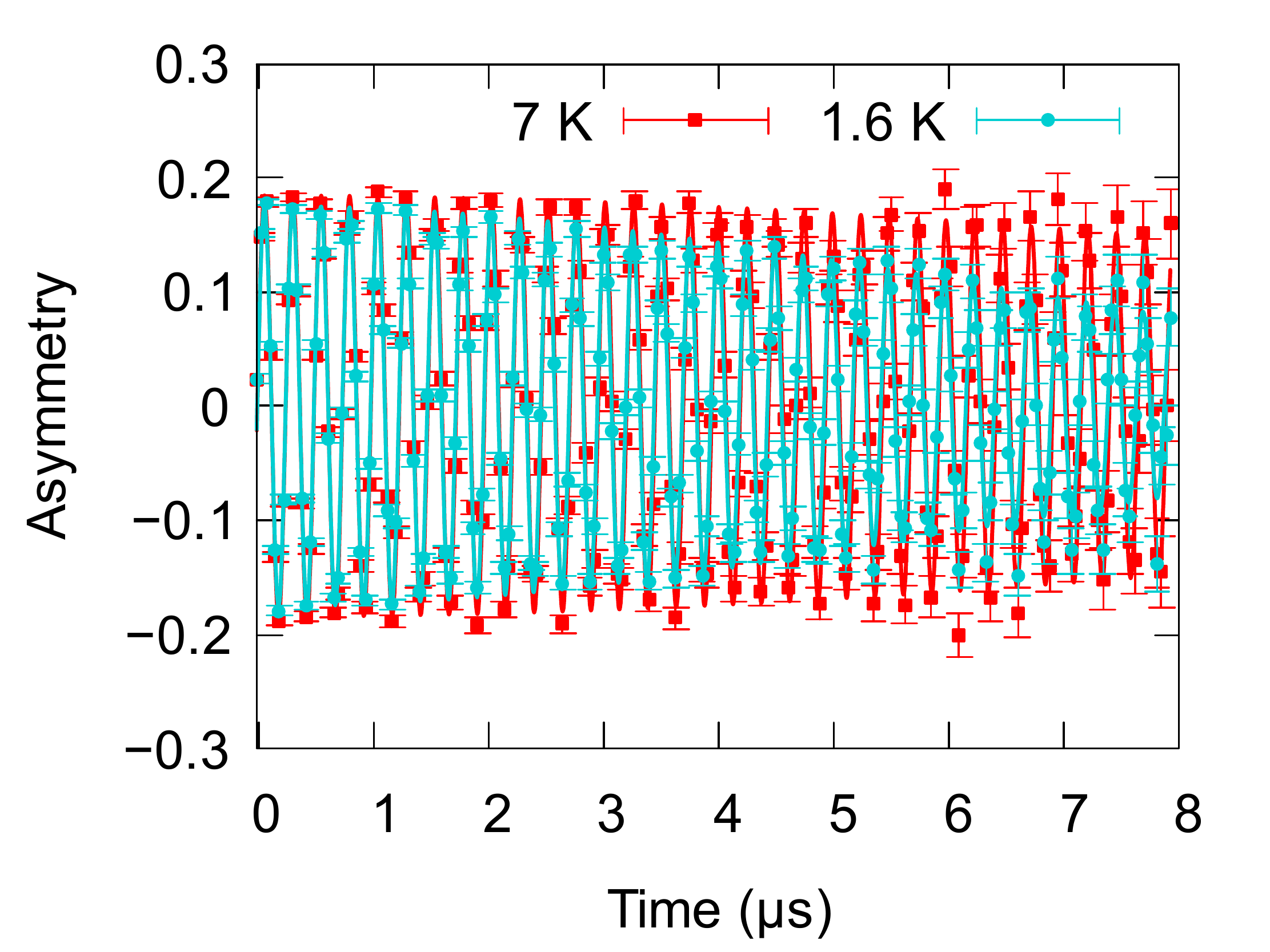}
\caption{Transverse-field $\mu$SR spectra for Sample A \rev{at} 30~mT \rev{and} 7~K (red squares) and 1.6~K (blue circles).
The oscillation of the asymmetry decays faster at 1.6~K than at 7~K due to the spatial distribution 
of the local magnetic field caused by flux-lattice formation.
The solid curves are results of the fittings with the Gaussian cosine function.}
\label{fig: PvsT_300Oe}
\end{figure}

\begin{figure}
\includegraphics[width=\linewidth]{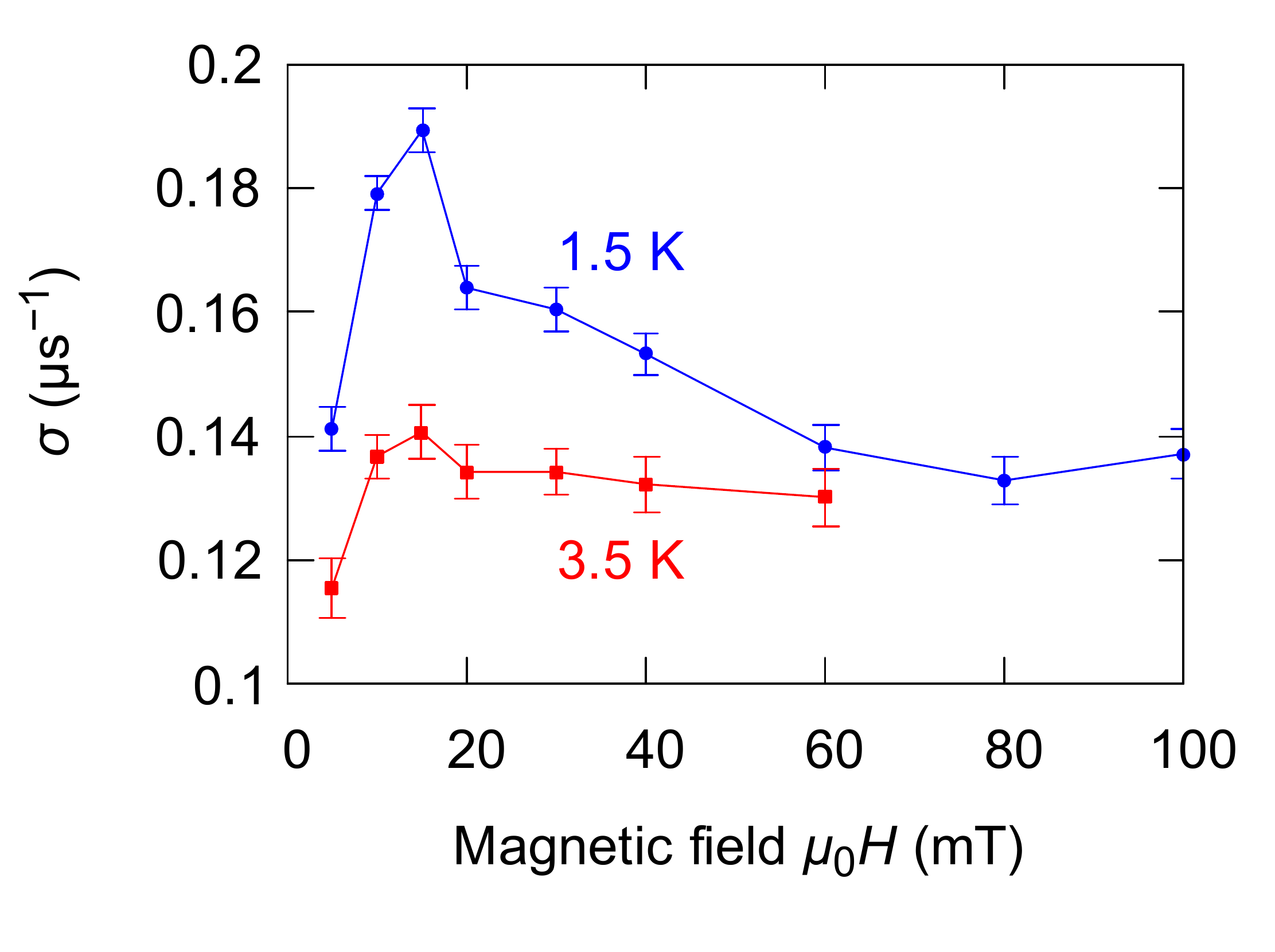}
\caption{Muon spin depolarization rate $\sigma$ of Sample A as a function 
of the applied magnetic field measured at 3.5~K (red squares) and 1.5~K (blue circles).
The enhancement of $\sigma$ below 20~mT and at 1.5~K are attributable to the superconductivity of Sn impurity 
with $H\sub{c}=30.9$~mT at $T=0$~K~\cite{Shaw1960SnHc}.}
\label{fig: PvsH}
\end{figure}

Next, we present in Fig.~\ref{fig: CPvsT} the electronic specific heat $C_e$ divided by temperature $T$ of Sample D at low temperature.
We observed a clear anomaly at around 0.85~K.
We fitted the temperature dependence using the Bardeen-Cooper-Schrieffer (BCS) theory~\cite{Tinkham2004Superconductivity} 
and the phononic contribution, which is already subtracted in Fig.~\ref{fig: CPvsT}.
We fixed $\Tc = 0.85$~K to satisfy the entropy balance.
We obtained the Sommerfeld coefficient $\gamma=5.67(3)$~mJ\,mol$^{-1}$K$^{-1}$, the Debye temperature $\Theta\sub{D}=119.9(2)$~K, 
and the volume fraction of 23\%.
This fact evidences the bulk nature of the 0.8~K phase.

\subsection{\boldmath$\mu$SR \rev{at} 30~mT}
\begin{figure}
\includegraphics[width=\linewidth]{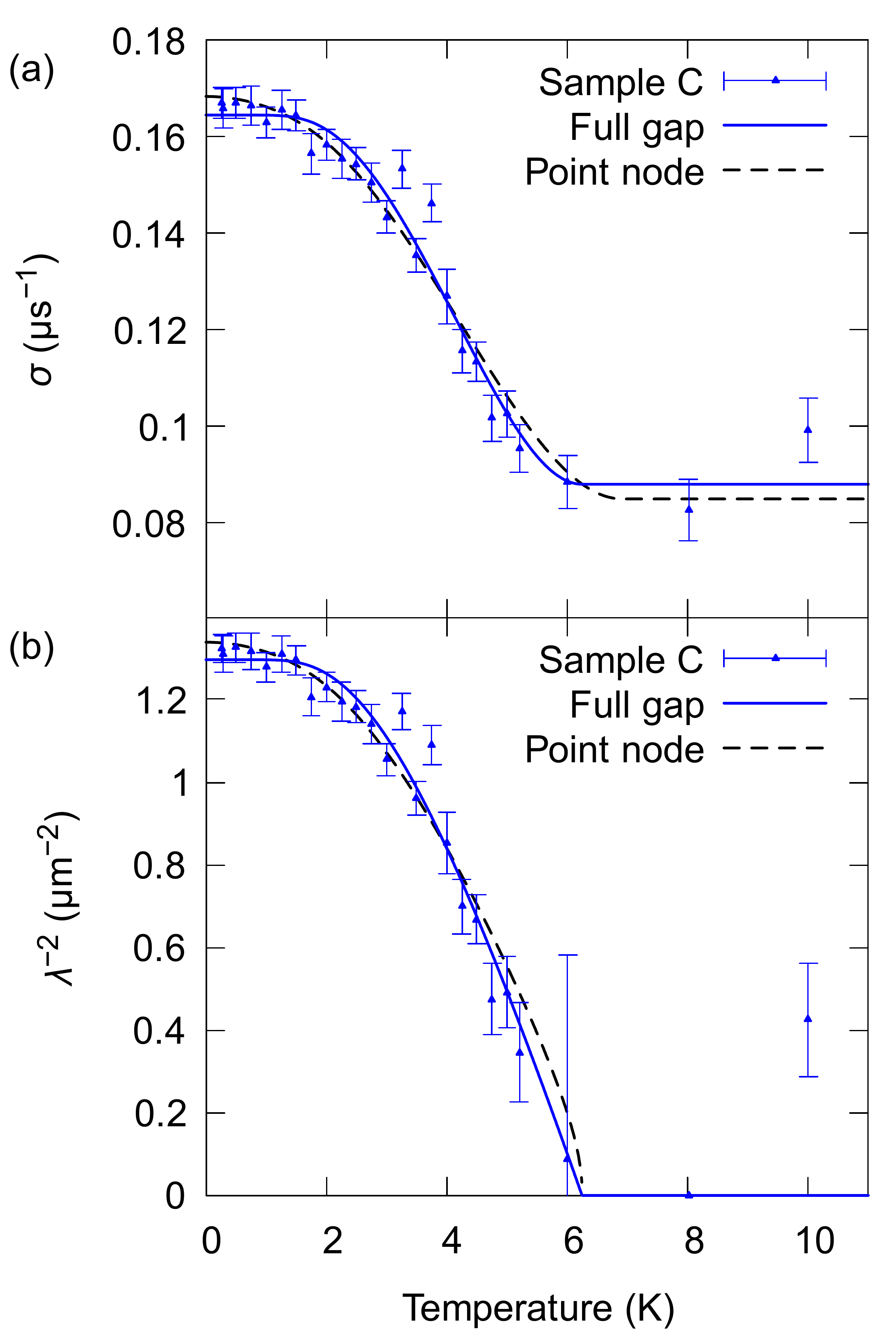}
\caption{Temperature dependence of (a) the muon spin depolarization rate $\sigma$ 
and (b) the penetration depth $\lambda$ for Sample C \rev{at} 30~mT\@.
The solid and dashed curves represent fitting results using the theoretical formulae 
for the fully gapped and point-nodal superconductors, respectively.
The fitting curve for the fully gapped state reproduces the experimental results better.
Note that the fitting gives a \Tc of 6.2~K, somewhat higher than the bulk \Tc of 5.4~K\@.}
\label{fig: gap_structure}
\end{figure}

Figure~\ref{fig: PvsT_300Oe} compares the $\mu$SR time spectra, recorded above and below \Tc,
measured under an applied field of 30~mT for Sample A\@.
The presence of the randomly oriented nuclear moments causes a weak relaxation of the $\mu$SR signal above \Tc.
The relaxation rate is enhanced below \Tc, which is caused by the formation of a flux-line lattice 
in the superconducting state, giving rise to an inhomogeneous magnetic field distribution.
Assuming the Gaussian distribution for the probability $n(B)$ that a muon stops at a position with a local flux density of $B$,
we fitted the data with the Gaussian cosine function 
\begin{equation}
A(t) = A(0)\exp\left(-\frac{\sigma^2t^2}{2}\right)\cos\left(2\pi\gamma_\mu{}B_0t+\phi\right),
\label{eqn: single_component}
\end{equation}
where $A(0)$ and $\phi$ are the asymmetry and phase at $t=0$, respectively, $\gamma_\mu$ is the gyromagnetic ratio of muon, 
and $B_0$ is the mean flux density inside the sample.
$\sigma$ changes from 0.084(6)~$\mu$s$^{-1}$ at 7~K to 0.160(4)~$\mu$s$^{-1}$ at 1.6~K\@.
This change in $\sigma$ is readily seen in the raw data, namely the faster damping at 1.6~K\@.
The change in the relaxation rate across \Tc indicates that a large portion of the muons stopped at the superconducting region of the sample.

We measured $\sigma$ as a function of the applied field at 1.5 K and 3.5 K (see Fig.~\ref{fig: PvsH}).
Each point was obtained by field cooling the sample from above \Tc to 1.7~K\@.
First, $\sigma$ strongly increases with increasing magnetic field until reaching a maximum at 15~mT 
and then above it continuously decreases up to the highest field (100~mT) investigated.
The observed field dependence of $\sigma$ implies that for a reliable determination 
of the penetration depth, the applied field must be just above the peak. 
Thus, we measured the temperature dependence of $\sigma$ at 30~mT\@. 

\begin{figure}
\includegraphics[width=\linewidth]{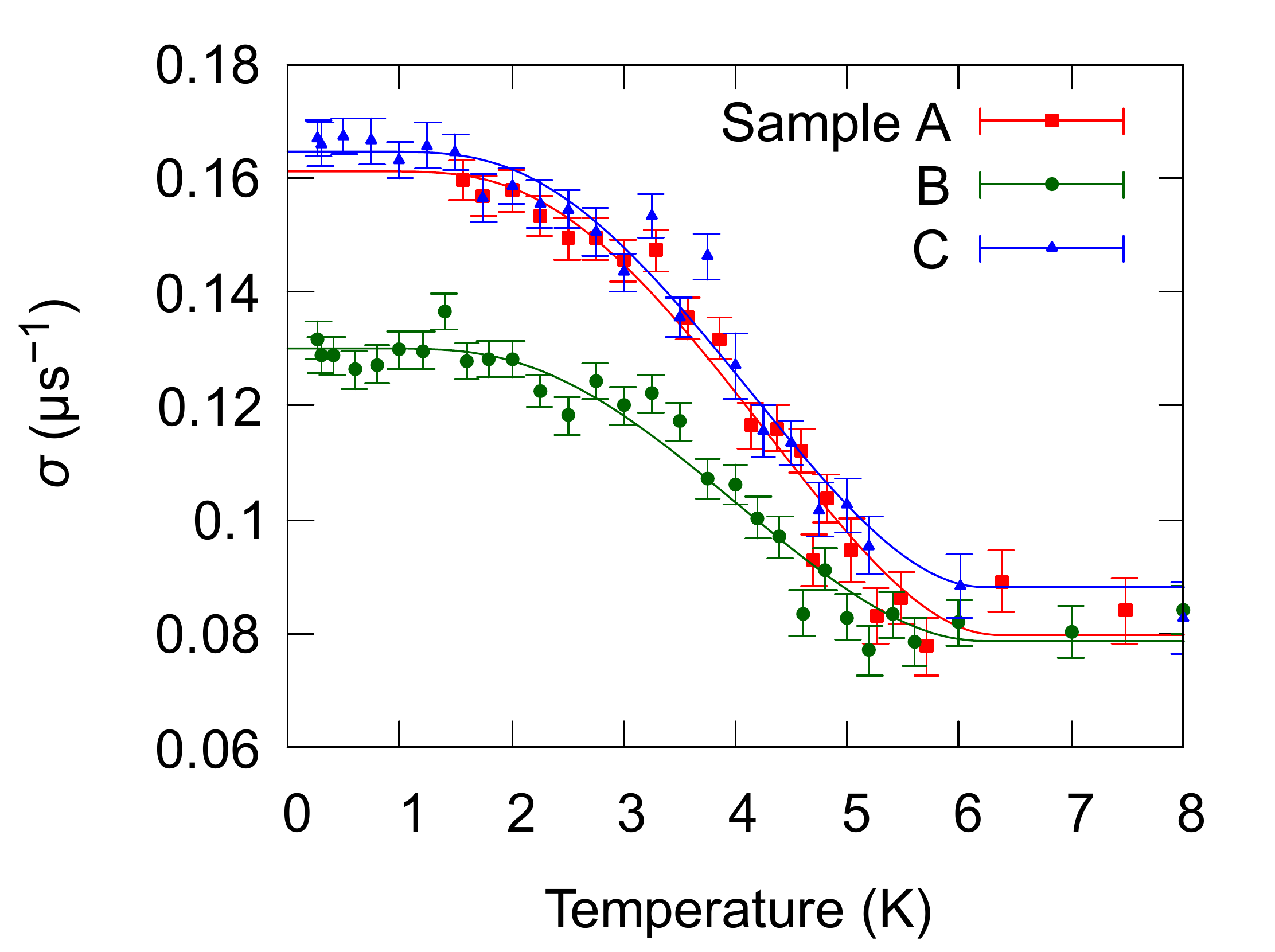}
\caption{Temperature dependence of the muon spin depolarization rate \rev{at} 30~mT\@.
The red squares, green circles, and blue triangles are the data for Samples A, B, and C, respectively.
The curve on each set of data represents the result of the \rev{fit} assuming a fully gapped superconducting state.}
\label{fig: SGMvsT_300Oe}
\end{figure}

% tables should appear as floats within the text
%
% Here is an example of the general form of a table:
% Fill in the caption in the braces of the \caption{} command. Put the label
% that you will use with \ref{} command in the braces of the \label{} command.
% Insert the column specifiers (l, r, c, d, etc.) in the empty braces of the
% \begin{tabular}{} command.
% The ruledtabular enviroment adds doubled rules to table and sets a
% reasonable default table settings.
% Use the table* environment to get a full-width table in two-column
% Add \usepackage{longtable} and the longtable (or longtable*}
% environment for nicely formatted long tables. Or use the the [H]
% placement option to break a long table (with less control than 
% in longtable).
\begin{table*}%[H] add [H] placement to break table across pages
%\color{red}
\caption{Transition temperature \Tc, penetration depth $\lambda(0)$, effective mass $m^*/m_e$, and $\Tc/\lambda(0)^{-2}$
of \SxSO extracted from $\mu$SR experiments, as resulting from fits with single and double Gaussian-cosine functions.}
\label{tbl: parameters}
\begin{ruledtabular}
\begin{tabular}{ccccccc}
Fit function & \multicolumn{3}{c}{Single Gaussian cosine: Eq.~(\ref{eqn: single_component})} & \multicolumn{3}{c}{Double Gaussian cosines: Eq.~(\ref{eqn: double_component})} \\
Sample & A & B & C & A & B & C \\
\hline
\Tc (K) & $6.3(2)$ & $6.2(2)$ & $6.2(2)$ & $5.6(3)$ & $6.05(7)$ & $6.04(5)$ \\
$\lambda(0)$ (nm) & $875(9)$ & $1018(11)$ & $878(10)$ & $316(23)$ & $340(11)$ & $322(15)$ \\
$m^*/m_e$ & $0.46$ & $0.61$ & $0.46$ & $0.053$ & $0.067$ & $0.060$ \\
$\Tc/\lambda(0)^{-2}$ (K$\,\mu$m$^2$) & $4.8$ & $6.4$ & $4.8$ & $0.56$ & $0.70$ & $0.62$
\end{tabular}
\end{ruledtabular}
\end{table*}

The temperature dependence of $\sigma$ for Sample C under an applied field of 30~mT is shown in Fig.~\ref{fig: gap_structure}.
An increase of $\sigma$ was observed below $\Tc=6$~K\@.
The relaxation rate is related to the London penetration depth $\lambda(T)$ \rev{via}
\begin{equation}
\sigma\sub{SC}(T) = \sqrt{0.00371}\times\frac{2\pi\gamma_\mu\Phi_0}{\lambda^2(T)},
\end{equation}
where $\sigma\sub{SC}(T)=\sqrt{\sigma^2(T)-\sigma\sub{N}^2}$ is the increase of the relaxation rate due to superconductivity, 
$\sigma\sub{N}$ is the relaxation rate in the normal state \rev{(}mostly attributable to the nuclear contribution\rev{)}, 
and $\Phi_0$ is the flux quantum~\cite{Brandt1988penetrationDepth}.
Moreover, the temperature dependence of $\lambda$ for the isotropic as well as 
anisotropic superconducting gaps can be calculated using the BCS theory:
\begin{equation}
\frac{\lambda^{-2}(T)}{\lambda^{-2}(0)} = 1+\frac{1}{2\pi}\int\int_0^\infty \frac{\partial f_{\bm{k}}}{\partial E_{\bm{k}}} \mathrm{d}\varepsilon\mathrm{d}\hat{\bm{k}},
\end{equation}
where $f_{\bm{k}}=\left[1+\exp\left(E_{\bm{k}}/(k\sub{B}T)\right)\right]^{-1}$ is the Fermi distribution function, 
$k\sub{B}$ is the Boltzmann constant, 
$E_{\bm{k}}=\sqrt{\varepsilon^2+\Delta_{\bm{k}}^2(T)}$ is the quasiparticle excitation energy of the superconducting state
with the kinetic energy $\varepsilon$ relative to the Fermi energy $\varepsilon\sub{F}$ and the superconducting 
gap $\Delta_{\bm{k}}(T)=\Delta_0(T)Y(\hat{\bm{k}})$~\cite{Tinkham2004Superconductivity}.
The temperature dependence of $\Delta_0$ is obtained by solving the gap equation~\cite{Hasselbach1993GapEquation}
\begin{equation}
\frac{1}{4\pi}\int\int_0^{\varepsilon\sub{c}} \frac{1}{E_{\bm{k}}}\tanh\left(\frac{E_{\bm{k}}}{2k\sub{B}T}\right) 
Y^2(\hat{\bm{k}}) \mathrm{d}\epsilon\mathrm{d}\hat{\bm{k}} = \mathrm{const.},
\end{equation}
where $\varepsilon\sub{c}$ is a cutoff energy.
The constant value in the right-hand side is numerically obtained by substituting $T=\Tc$ and $\Delta_0=0$.
We simply set the cutoff energy and the Fermi energy as $\varepsilon\sub{c}=100k\sub{B}T\sub{c}$ and $\varepsilon\sub{F}=2000k\sub{B}T\sub{c}$.
These values are compared to the Debye temperature $\varepsilon\sub{D}=35k\sub{B}\Tc$ 
measured by \ce{^{119}Sn}-M\"{o}ssbauer spectroscopy~\cite{Ikeda2019Moessbauer} and 
$\varepsilon\sub{F}=2000k\sub{B}\Tc$ estimated from the band structure calculation~\cite{Ikeda2018bandstructure}.

We calculated the temperature dependence of the relaxation rate assuming two superconducting gap structures on a spherical Fermi surface:
$Y(\hat{\bm{k}})\equiv1$ for a fully gapped and $Y(\hat{\bm{k}})=\sqrt{\hat{k}_x^2+\hat{k}_y^2}$ for a point-nodal states.
While $\sigma$ saturates at low temperatures for a fully gapped state (the solid curve in Fig.~\ref{fig: gap_structure}), 
$\sigma$ continues to increase as lowering temperature for a point-nodal state (the dashed curve).
Both fitting curves reasonably match the experimental result within the experimental error,
but the root mean square error is smaller for the fully gapped state (1.2758) than for the point-nodal state (1.3933).
Although it is theoretically suggested that the superconductivities with a full gap and point nodes have similar 
transition temperatures~\cite{Kawakami2018TopoSC}, this fitting analysis suggests that the fully gapped superconductivity 
such as the ordinary $s$-wave superconductivity or the topological superconductivity 
resembling the BW state is more likely to be realized in \SxSO.

Figure~\ref{fig: SGMvsT_300Oe} compares $\sigma(T)$ of all samples measured \rev{at} 30~mT\@.
\rev{We fitted $\sigma(T)$ assuming a fully gapped superconducting wavefunction.
The resulting values of \Tc and $\lambda(0)$ are summarized in Table~\ref{tbl: parameters}.}
For all samples, we obtained higher \Tc than those in the magnetic measurements in Fig.~\ref{fig: Susceptibility},
probably because $\mu$SR detects the superconducting transition of a small part of the sample with slightly higher \Tc.
Using the coherence length of $\xi(0)=27$~nm evaluated from the upper critical field~\cite{oudah2016superconductivity}, 
the Ginzburg-Landau parameter $\kappa$ is estimated to be $\kappa=\lambda(0)/\xi(0)=32$--38, suggesting a strongly type-II superconductivity.

$\lambda^{-2}$ is related to the superfluid density $n\sub{S}$ and effective mass $m^*$ through the equation
\begin{equation}
\lambda^{-2}=\frac{\mu_0n\sub{S}e^2}{m^*}\times\frac{1}{1+\xi/l},
\end{equation}
where $\mu_0$ is the magnetic permeability in vacuum, 
$e$ is the elementary charge, and $l$ denotes the mean free path~\cite{Tinkham2004Superconductivity}.
Since the density of states at the Fermi energy $D(\varepsilon\sub{F})$ is evaluated to be
$D(\varepsilon\sub{F})=1.203(7)$~eV$^{-1}$ per unit cell per spin from $\gamma$, the superfluid density is estimated to be 
$n\sub{S} = \Delta_0 \times D(\varepsilon\sub{F}) \simeq 1.76k\sub{B}T\sub{c}D(\varepsilon\sub{F}) =  1.7\times10^{25}$~m$^{-3}$.
Therefore, the effective mass is calculated to be
$m^* = \mu_0n\sub{S}e^2\lambda^2/(1+\xi/l) < \mu_0n\sub{S}e^2\lambda^2$ in the clean limit.
\rev{The effective mass normalized by the rest mass of the electron $m^*/m_e$ for each sample is listed in Table~\ref{tbl: parameters}.}
If the samples are in the dirty limit, $m^*$ is further reduced.

\begin{figure}
\includegraphics[width=\linewidth]{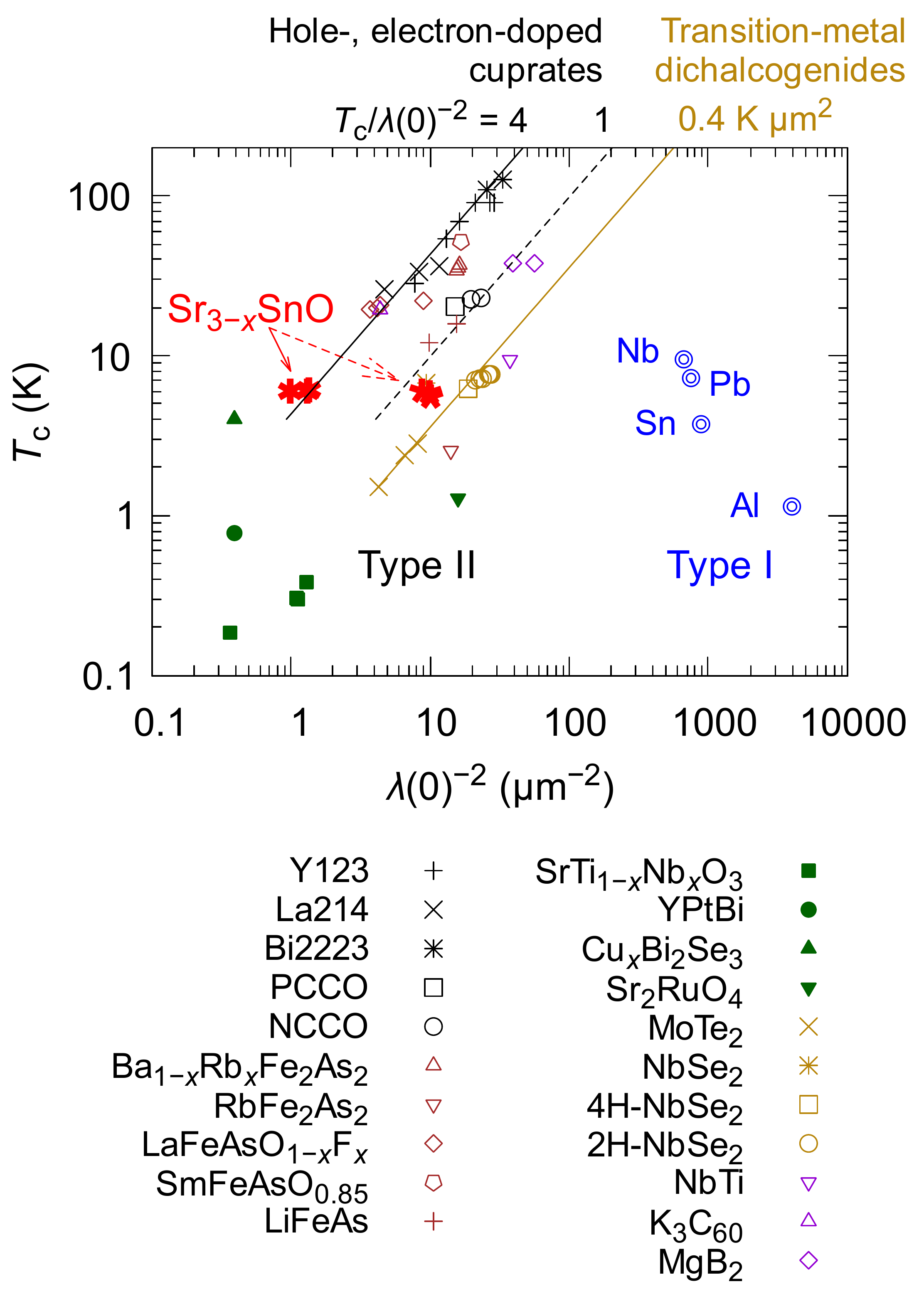}
\caption{Plot of \Tc vs $\lambda(0)^{-2}$ (Uemura plot) for cuprate (black)~\cite{Kim2012UemuraPlot}, 
iron pnictide (brown)~\cite{Khasanov2008SmFeAsO, Luetkens2009LaFeAsO1-xFx, 
Pratt2009LiFeAs, Shermadini2010RbFe2As2, Guguchia2011Ba1-xRbxFe2As2}, 
low-carrier or unconventional (green)~\cite{Riseman1998Sr2RuO4, Bay2014YPtBi, Collignon2017SrTiO3, Krieger2017CuxBi2Se3}, 
transition-metal-dichalcogenide (orange)~\cite{Le1991NbSe2, Guguchia2017MoTe2, vonRohr2019TMD}
elemental $s$-wave (blue)~\cite{Kittel2005SolidStatePhysics}, 
and compound isotropic (purple)~\cite{Uemura1991KC60, Gauzzi2000NbTi, Chen2001MgB2, Carrington2003MgB2} superconductors.
The solid and dashed lines represent the relations for the hole- and electron-doped 
cuprates and the transition metal dichalcogenides, namely $\Tc/\lambda(0)=4$, 1, and 0.4~K\,$\mu$m$^2$, 
respectively~\cite{Uemura1989UemuraPlot, Shengelaya2005Sr0.9La0.1CuO2}.
The large six- and five-spoked asterisks represent the data for \SxSO obtained 
from analyses with single and double Gaussian cosine functions, respectively.}
\label{fig: Uemura plot}
\end{figure}

It has been known that \Tc and $\lambda(0)^{-2}$ of various materials exhibit a scaling behavior with the ratio $\Tc/\lambda(0)^{-2}$ 
depending on the class of superconductors, as summarized in the Uemura plot~\cite{Uemura1989UemuraPlot, Uemura1991UemuraPlot}.
For some class of superconductors, \Tc is very high despite a small $n\sub{S}$.
For example, hole-doped cuprate and iron-based superconductors satisfy 
the relation $\Tc/\lambda(0)^{-2}\simeq4$~K$\,\mu$m$^2$~\cite{Uemura1989UemuraPlot}, 
whereas \rev{elementary} superconductors exhibit much smaller $\Tc/\lambda(0)^{-2}$ 
of less than 0.02~K$\,\mu$m$^2$~\cite{Kittel2005SolidStatePhysics}.
For the low-carrier superconductors \ce{SrTi_{1-x}Nb_xO3}~\cite{Collignon2017SrTiO3} and YPtBi~\cite{Bay2014YPtBi}, 
the ratio is $\Tc/\lambda(0)^{-2}=0.267$--0.496 and 2.0~K$\,\mu$m$^2$, respectively.
Interestingly, the obtained values of \SxSO ($\Tc=6$~K and $\lambda(0)=0.9$--1~$\mu$m) give 
\rev{a} $\Tc/\lambda(0)^{-2}$ ratio close to those of high-temperature or unconventional superconductors 
as presented in \rev{Table~\ref{tbl: parameters} and Fig.~\ref{fig: Uemura plot}.}
These values \rev{of $\Tc/\lambda(0)^{-2}$} are between those of the high-temperature superconductors and of the topological superconductor 
Cu-intercalated \ce{Bi2Se3}, for which $\Tc/\lambda(0)^{-2} = 10$~K$\,\mu$m$^2$ has been reported~\cite{Krieger2017CuxBi2Se3}.
The relatively high \Tc for a small number of carriers implies an unconventional superconducting mechanism in \SxSO.

We also performed analyses to take into account the volume fraction of the superconductivity.
To do this, we fitted the time-dependent asymmetry with two Gaussian cosine functions
\begin{equation}
A\left[\alpha\exp\left(-\frac{\sigma^2t^2}{2}\right)+\exp\left(-\frac{\sigma\sub{N}^2t^2}{2}\right)\right]\cos\left(2\pi\gamma_\mu{}B_0t+\phi\right),
\label{eqn: double_component}
\end{equation}
where $\sigma\sub{N}$ and $\phi$ are fixed to the values above 6~K and $\alpha$ is fixed to 0.1 for Samples A and B and to 0.04 for Sample C
corresponding to the volume fraction of 10 and 4\% estimated from the susceptibility measurement in Fig.~\ref{fig: Susceptibility}, respectively.
\rev{The results of the two-component fits are summarized in Table~\ref{tbl: parameters}.}
The $\Tc/\lambda^{-2}$ ratios are comparable to those of the low-carrier superconductors.
\rev{This fact again indicates the relatively high \Tc for a small number of carriers in \SxSO.}
Thus, the qualitative conclusion is still valid even if the phase separation is taken into account.

\begin{figure}
\includegraphics[width=\linewidth]{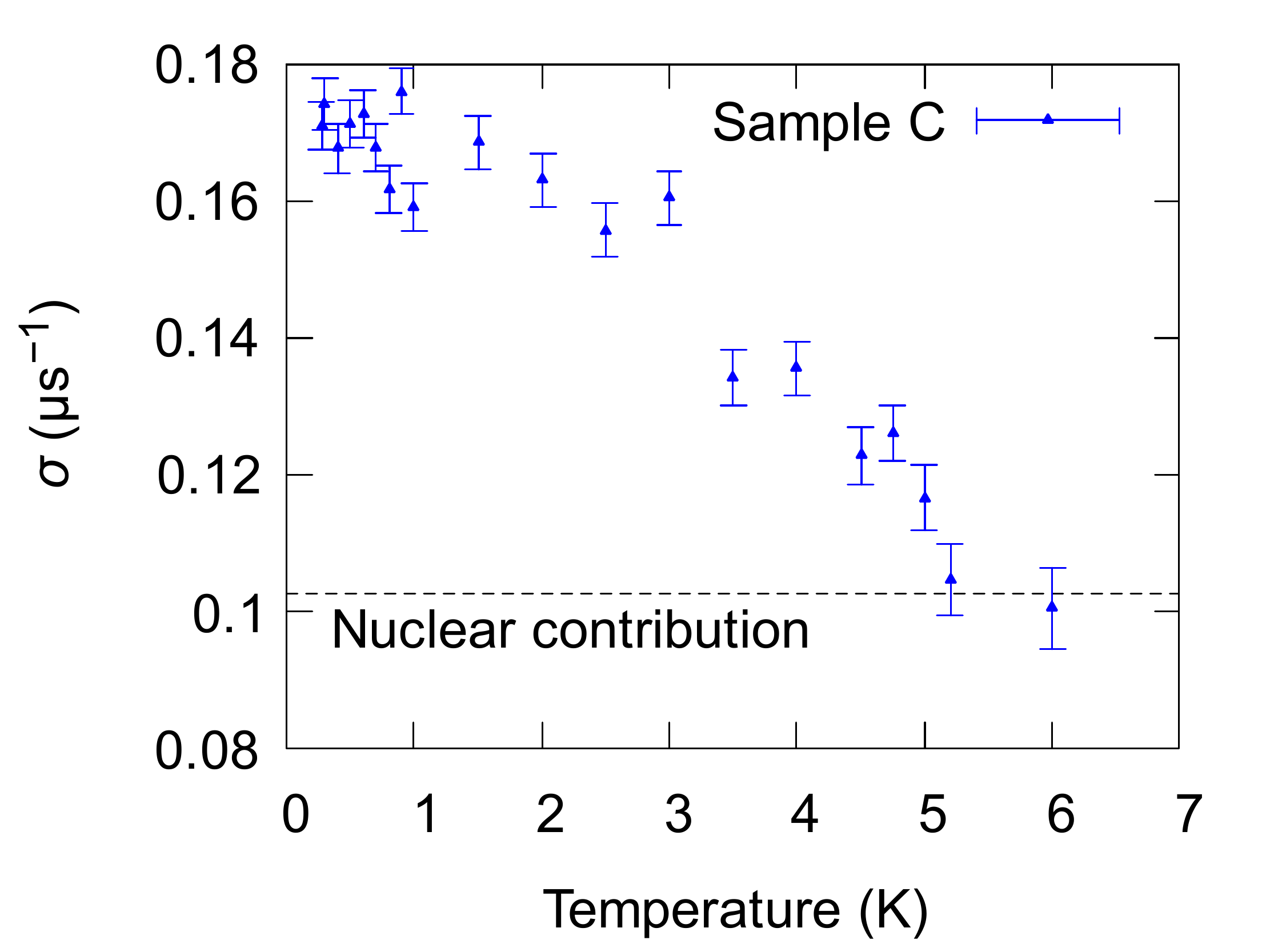}
\caption{Temperature-dependent relaxation rate of Sample C \rev{at} 5~mT\@.
The large increase at 3~K probably originates from the superconductivity of Sn impurity.}
\label{fig: SGMvsT_50Oe}
\end{figure}

\subsection{\boldmath$\mu$SR \rev{at} 5 and 0~mT}
Figure~\ref{fig: SGMvsT_50Oe} shows the relaxation rate of Sample C \rev{at} 5~mT\@.
The jump at 3.0~K probably originates from the superconductivity of Sn.
We did not detect the increase of the relaxation rate at 0.8~K, 
probably because the field modulation caused by the 0.8-K phase is too small \rev{at} 5~mT
even though the superconductivity originates from the bulk.
The penetration depth of the 5.4-K phase \rev{at} 5~mT is calculated to be $8.9(2)\times10^{-7}$~m.
This value is consistent with the one \rev{at} 30~mT within the error.
Since the magnetic field of 20~mT is reported to completely suppress the 0.8-K phase~\cite{Oudah2019SrDeficiency}, 
the upper critical field of the 0.8-K phase may not be large enough for \ce{\mu}SR.
Measurement of the field dependence of the relaxation rate below 0.8~K remains as a technical challenge in a future work.

\begin{figure}
\includegraphics[width=\linewidth]{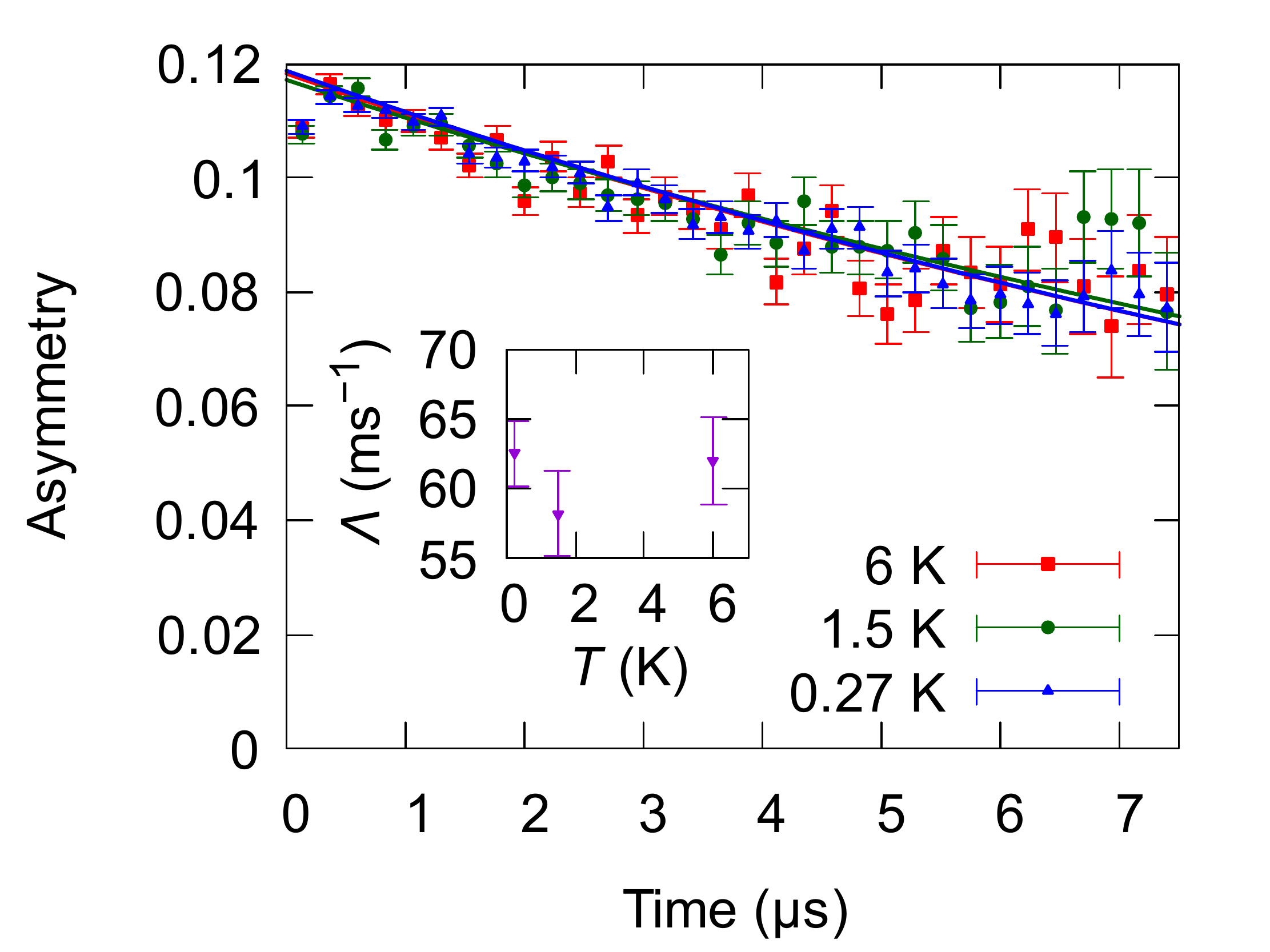}
\caption{Zero-field $\mu$SR spectra of Sample C, recorded at 6~K (red squares), 1.5~K (green circles), and 0.27~K (blue triangles).
The solid curves represent the fitting results with the exponential function.
Since the signals at different temperatures are on top of each other within the error bars, 
the time-reversal symmetry is preserved in the superconducting state of \SxSO.
Indeed, the temperature dependence of the relaxation rate $\Lambda$ is quite weak as shown in the inset.}
\label{fig: PvsT_0Oe}
\end{figure}

Finally, we present in Fig.~\ref{fig: PvsT_0Oe} the time dependence of the asymmetry \rev{at} zero field to test 
a possible spontaneous time-reversal-symmetry breaking.
The data were fitted with the exponential function $A\exp(-\Lambda t)$, and
we obtained $\Lambda$ to be 62(3)~ms$^{-1}$ at 6~K, 58(3)~ms$^{-1}$ at 1.5~K, and 63(2)~ms$^{-1}$ at 0.27~K\@.
Thus, temperature dependence of $\Lambda$ is quite weak, as shown in the inset of Fig.~\ref{fig: PvsT_0Oe}.
The possible maximum spontaneous flux density due to superconductivity is evaluated to be 
$\left(\Lambda\sub{0.27 K}-\Lambda\sub{6 K}\right)/\left(2\pi\gamma_\mu\right)=7$~$\mu$T\@.
This maximum value is seven-times smaller than that observed in \ce{Sr2RuO4} (50~$\mu$T)~\cite{Luke1998Sr2RuO4}.
Thus, we conclude that the time-reversal symmetry is very likely to be preserved in the superconducting state of \SxSO.
The preserved time-reversal symmetry is consistent with the theoretically proposed 
topological phase as well as the ordinary $s$-wave superconductivity.

% Surround table environment with turnpage environment for landscape
% table
% \begin{turnpage}
% \begin{table}
% \caption{\label{}}
% \begin{ruledtabular}
% \begin{tabular}{}
% \end{tabular}
% \end{ruledtabular}
% \end{table}
% \end{turnpage}

\section{Conclusion\label{conclusion}}
In this paper, we report the first $\mu$SR \rev{investigation of} the antiperovskite oxide superconductor \SxSO.
The temperature dependence of the muon spin depolarization rate, measured 
in the vortex state of \SxSO at low temperatures, suggests a fully gapped superconductivity.
The London penetration depth is estimated to be around 320--1020~nm \rev{(depending on samples and type of analysis)},
and the ratio $\Tc/\lambda(0)^{-2}$ is similar to those of high-temperature superconductors or unusual low-carrier superconductors.
This fact may hint \rev{at} unconventional superconducting mechanism in \SxSO.
In addition, we did not detect any sign \rev{of broken} time-reversal symmetry.
These features are consistent with the theoretically proposed topological phase, 
namely superconductivity similar to the B phase of \ce{^3He} or with the conventional $s$-wave superconductivity.
%\vspace{6\baselineskip}

% Specify following sections are appendices. Use \appendix* if there
% only one appendix.
%\appendix
%\section{}

% If you have acknowledgments, this puts in the proper section head.
\begin{acknowledgments}
We thank Y. J. Uemura for \rev{insightful recommendations and suggestions and for helping us in planning the $\mu$SR experiments}.
We are grateful to C. Baines for the technical support and discussions.
We also thank W. Higemoto and V. Grinenko for useful discussions.
We acknowledge the Research Center for Low Temperature and Materials Sciences in Kyoto University for the supply of liquid He.
This work was partially supported by Japan Society for the Promotion of Science (JSPS) KAKENHI 
No.\ JP15H05851, JP15H05852, JP15K21717 (Topological Materials Science), JP17H04848, and JP17J07577, 
and by the JSPS Core-to-Core Program (A. Advanced Research Network), 
as well as by the Izumi Science and Technology Foundation (Grant No.\ H28-J-146).
A.I. is supported by the JSPS Research Fellowship.
M.O. is supported by the Max Planck-UBC-UTokyo Center for Quantum Materials.
\end{acknowledgments}

% Create the reference section using BibTeX:
%\bibliography{../antiperovskite}

%apsrev4-2.bst 2019-01-14 (MD) hand-edited version of apsrev4-1.bst
%Control: key (0)
%Control: author (8) initials jnrlst
%Control: editor formatted (1) identically to author
%Control: production of article title (0) allowed
%Control: page (0) single
%Control: year (1) truncated
%Control: production of eprint (0) enabled
%
\end{document}